\documentclass[preprint,12pt]{aastex}
\usepackage{graphics}

\shorttitle{Outer Edge of Saturn's B-ring}
\shortauthors{Spitale et al.}

\begin{document}

\title{Free Unstable Modes and Massive Bodies in Saturn's Outer B Ring}

\author{J. N. Spitale}
\affil{CICLOPS, Space Science Institute}
\affil{4750 Walnut St. Ste 20, Boulder, CO 80301}
\email{joes@ciclops.org}

\author{C. C. Porco}
\affil{CICLOPS, Space Science Institute}
\affil{4750 Walnut St. Ste 205, Boulder, CO 80301}

\pagebreak

\noindent \textbf{Running Head:Outer Edge of Saturn's B-ring}  \\ \\

\noindent \textbf{Editorial correspondence:}

\noindent Joe Spitale \\
\noindent 2719 E. Mabel St. \\
\noindent Tucson, AZ 85716 \\

\noindent email: joes@ciclops.org \\
\noindent phone: (520)207-8782 \\

\pagebreak
\doublespace

\section*{Abstract} 
Voyager images and Cassini occultation data have previously shown that the behavior of the outer edge of Saturn's massive B ring is determined only in part by a static response to the 2:1 inner Lindblad resonance with Mimas. In  Cassini images of this region, we find, in addition to the expected wavenumber-2 forced distortion, evidence for unforced self-excited wavenumber-3, wavenumber-2, and wavenumber-1 normal modes.  These are the first observations to suggest substantial wave amplification in Saturn's broad rings.  Moreover, the presence of these free modes strongly implicates viscous overstability as their underlying cause and, by inference, the cause for most if not all of the unforced structures throughout the high-mass-density B ring and in other high-mass-density regions in Saturn's rings.  Analysis of each of the inferred waves reveals a consistent lower bound on the average surface mass density of $\sim$ 44 g cm$^{-2}$ for the outer 250 km of the ring, though the true surface density could be as high as 100 g cm$^{-2}$ or higher.  Interference between the forced and free wavenumber-2 modes yields a total wavenumber-2 pattern that varies in amplitude and orientation with a characteristic period of $\sim$ 5.5 years.  We also find localized disturbances, including 3.5-km-tall vertical structures, that provide circumstantial evidence for embedded massive bodies in the Mimas resonance zone.  The presence of such bodies is supported by the presence of a shadow-casting moonlet $\sim$ 0.3 km wide near the ring's edge.

\noindent \emph{Subject headings:} planets:rings

\section{Introduction} 
The outer edge of Saturn's B-ring is confined by the strongest resonance in Saturn's ring system: the 2:1 inner Lindblad resonance (ILR) with Mimas \citep{Goldreich1978}.  That resonance is expected to impose a variation in the radial position of the edge with azimuthal wavenumber $m=2$ (i.e., a Saturn-centered ellipse), with one minimum locked to Mimas \citep{Goldreich1982}.  Voyager observations were consistent with those predictions \citep{Porco1984}, implying a radial amplitude of about 75 km, though relatively large residuals indicated additional unresolved structure.  A preliminary analysis of images taken early in the Cassini mission confirmed the existence of the Saturn-centered ellipse, but found Mimas lagging the radial minimum by about 28$^\circ$, suggesting either a static lag between the ring response and the direction to the satellite, or a libration of the periapse direction relative to the satellite, or both.  We also reported higher-wavenumber radial oscillations and additional time variations that were difficult to characterize, though an $m=3$ pattern did seem to explain part of that variation \citep{Spitale2006}.  Here, we use Cassini imaging data sets spanning about 4 years to analyze the B-ring outer edge in greater detail than previously attainable, with the objective of understanding its complicated shape.  

For the kinematical modeling in this paper, we examined 18 data sets spanning the interval from day 2005-174 to day 2009-207, consisting of a total of nearly 2000 Cassini Narrow-Angle Camera (NAC) images (see Table \ref{tbl:data}), with pixel scales about 10 km or better (Fig. \ref{fig:maps}), employing the techniques and types of image sequences (azimuthal scans and ansa-staring movies) described in our earlier work \citep{Spitale2009}.  We also examined several smaller data sets taken near 2009 Saturn equinox, when the sun elevation was low enough for shadows cast by out-of-plane material to be detectable.

\section{Kinematical Modeling}  \label{sec:modeling}

Fig. \ref{fig:maps} shows map-projected mosaics in radius vs. longitude for each data set used in the kinematical modeling.  An additional data set, set 64, is shown as well because it spanned a significant range of co-rotating longitudes, though it was excluded from the kinematical modeling because most of the images did not contain the fiducial feature (feature number 13 from \citet{French1993}) used for all of the other data sets to determine radial positions.  The expected 2-lobed shape is apparent in many of the mosaics in Fig. \ref{fig:maps}, with additional higher wavenumber variations superimposed.  However, the amplitude of the $m=2$ pattern varies, and the pattern is absent, or barely discernible in some data sets (e.g., sets 6, 8, 9, and 23).  

As in almost all previous analyses of planetary rings, including our recent study of the outer edge of the A-ring \citep{Spitale2009}, we modeled the ring edge using a sinusoidal model of a streamline to describe each mode whose presence we investigated.  After some experimentation with linear combinations of various low-wavenumber modes, we found the best-fitting model to be one comprising 4 normal modes: an $m=1$ moving at a rate appropriate for free precession near the outer B-ring edge, an $m=2$ moving with a pattern speed $\Omega_p$ expected for forcing by the Mimas 2:1 ILR (i.e., Mimas' average mean motion during the period of the investigation, see Sec. \ref{sec:m=2}), a second $m=2$ moving slightly faster than Mimas, and an $m=3$ mode.  

Parameters for the adopted solution are given in Table \ref{tbl:elem_m=1224}, and Fig. \ref{fig:plots} plots each component.  The large $\chi^2$/DOF value indicates that there is significant additional structure to be characterized that we were unable to model with simple normal modes.  However, note that the largest deviations from the four-component model (colored red and blue in the figure) tend to comprise radial excursions of narrow ($\sim$30$^\circ$) azimuthal extent, which are reminiscent of the ``spikes'' pointed out in our recent study of the A-ring  outer edge \citep{Spitale2009}.  For reasons discussed below, those spikes were considered to be independent of the edge modes, and were thus excluded from the kinematical fits.  Therefore, the quoted statistics do not include those large excursions.

\section{$\mathbf{m=1}$ Disturbance}
We interpret the presence of an $m=1$ normal mode at the outer B-ring edge as an unforced self-excited unstable normal mode or standing density wave.  Dense rings, like the outer B ring, behave like non-Newtonian fluids and the close packing of the particles can lead to collective behaviors similar to granular flow.  Under these circumstances density waves are expected to become viscously overstable; in the presence of feedback, as occurs upon double reflection within a resonant cavity, they become non-linear \citep{Borderies1985}.  Very short wavelength structures -- of order $\sim$ 150 to 400 m -- have been observed in Cassini imaging \citep{Porco2005} and occultation data \citep{Colwell2007} within the high optical depth ($\tau \sim 1.5$) B-ring, and have been interpreted as spontaneously formed, axisymmetric, non-linear waves or ``viscous overstabilities'' in both cases.  These axisymmetric modes are just one form of overstability; non-axisymmetric modes are possible as well and are believed to be responsible for the multi-mode shapes of the narrow rings of both Saturn and Uranus \citep{Porco1990}.  The density waves with the longest wavelengths are those with $m=1$.  However, without feedback and significant amplification, unforced density wave trains, especially of such long wavelength, will damp.  

The existence of an $m=1$ disturbance at the outer B-ring edge whose radial amplitude ($ae \simeq 21$ km) is comparable to that of the $m=2$ resonant perturbation ($ae \simeq 35$ km; see below) argues that sufficient feedback and hence amplification must exist for this mode to be observable.  Amplification requires a resonant cavity within which the wave can reflect and grow.  The outer B-ring edge, at $a_0 = 117568(4)$ km, provides a sharp outer boundary for a resonant cavity; the inner boundary is formed by the mode's ILR whose location is determined by solving $(m-1)n(a) + \dot\varpi(a) - \Omega_p = 0$ for $a$, where $n(a)$ and $\dot\varpi(a)$ are the mean motion and apsidal precession rate at semimajor axis $a$ respectively.  For the $m=1$ speed given in Table \ref{tbl:elem_m=1224}, the ILR location is $a_l$ = 117314.92(8) km.  The resonant cavity is consequently 253(4) km wide.

If the wavelength of an $m=1$ disturbance propagating away from a Lindblad resonance is known, we can estimate the surface mass density required to keep the wave from shearing due to differential precession.  In the general formulation for the behavior of linear density waves driven at a resonance in a ring of surface mass density $\Sigma$, the first wavelength $\Lambda$ is given by \citet{Goldreich1978}:
\begin{equation}
	\frac{2\pi}{\Lambda} = D\frac{x}{2\pi G \Sigma}, %
                                                                 \label{eq:cp_a}
\end{equation}
where $x$, the fractional distance from an ILR location $a_l$ is equal to $(a - a_l)/a_l$ and  \citep{Marley1993}:
\begin{equation}
	D = n^2 \left\{\left[3 - \frac{9}{2} J_2 %
                   \left(\frac{R_p}{a_l}\right)^2\right] (m - 1) %
                   + \frac{21}{2} J_2 \left(\frac{R_p}{a_l}\right)^2\right\}. %
                                                                 \label{eq:cp_x}
\end{equation}
\noindent Also, $G$ is the universal constant of gravitation, $J_2$ is Saturn's second zonal gravity harmonic and $R_p$ is Saturn's reference radius.  If the ring edge at $a=a_0$ provides a fixed outer boundary to the cavity, imposing a node at that radial location then, for the case in which no radial nodes exist within the cavity, 1/2 of the first wavelength of the density wave must fit exactly within the cavity.  Consequently, $\Upsilon\Lambda$ must equal the cavity width $a_0 - a_l$, where $\Upsilon$ = 1/2.  However, the ring edge cannot be fixed because it is not circular, its shape being perturbed by these non-axisymmetric density waves, including the one created by the Mimas 2:1 resonance.  Therefore the condition for a node at the ring edge does not exist and the cavity may actually support a longer wave, in which case $\Upsilon$ would be less than 1/2.  We solve for the wavelength at the location where $a_0 - a_l = \Upsilon\Lambda$, taking $n^2 \sim G M_p/a_l^3$, to obtain:
\begin{equation}
	\frac{\Lambda^2}{\Sigma} = \frac{4\pi^2 a_l^4} %
	       {\Upsilon M_p\left\{\left[3 - \frac{9}{2} J_2 %
                   \left(\frac{R_p}{a_l}\right)^2\right] (m - 1) %
                   + \frac{21}{2} J_2 \left(\frac{R_p}{a_l}\right)^2\right\}}, %
                                                                 \label{eq:cp_b}
\end{equation}
where $M_p$ is Saturn's mass.  Setting $\Upsilon\Lambda$ = 253(4) km yields a lower bound on the average surface mass density across this region of $\Sigma$ = 43.5(7) g cm$^{-2}$, for $\Upsilon$ = 1/2.  The Cassini Radio Science optical depth for this putative resonant cavity is $\tau \sim$ 1.2 for the outermost $\sim 100$ km, and $\tau \sim$ 0.6 interior to that (Fig. 13.15 in \citet{Colwell2009}).  Regions with $\tau \sim$ 0.8 in the A-ring have been found to have $\Sigma \sim$ 40 g cm$^{-2}$ \citep{Tiscareno2007,Tiscareno2009DPS}, comparable to our lower bound for the B ring.  If $\Upsilon$ is equal to, say, 1/4 because the edge is not fixed, Eq. \ref{eq:cp_b} yields a surface density of 87.0(6) g cm$^{-2}$.

As support for the presence of an $m=1$ standing wave in the outer few hundred km of the B-ring, we note that this is one of the few places in Saturn's rings where multiple occultation profiles diverge significantly from one another (see Fig. 11 of \citet{Nicholson2000}), as would be expected if a non-axisymmetric structure like an $m=1$ standing wave were present.

\section{$\mathbf{m=3}$ Pattern}  \label{sec:m=3}
In a previous analysis of the inner Cassini Division \citep{Spitale2006}, we noted that a wavenumber-3 pattern moving at $\sim$ 505.5 $^\circ$/day provided a good description for the residuals in the data sets that we have labeled 3 and 5 in this work.  Here, we find that such a mode, moving at $\sim$ 507.7$^\circ$/day, is consistent with the entire set of observations, as shown in Table \ref{tbl:elem_m=1224} and Fig. \ref{fig:plots}.  Again, we interpret this pattern as an unstable free oscillation, whose best-fit pattern speed implies an ILR at 117544.2(4) km.  The resonant cavity within which the $m=3$ mode resides is therefore $117568(4) - 117544.2(4) = 24(5)$ km wide.  With $\Upsilon$ = 1/2, we calculate from Eq. \ref{eq:cp_b} a lower bound for the surface mass density of $\Sigma = 51(11)$ g cm$^{-2}$; for $\Upsilon$ = 1/4, the surface density is 100(11) g cm$^{-2}$.  Both values are consistent with those derived for the $m=1$ free mode.

\section{$\mathbf{m=2}$ Pattern: Forced and Free} \label{sec:m=2}
To more clearly illustrate the time variation in the $m=2$ shape, we show independent fits to the total $m=2$ pattern for each data set in Fig. \ref{fig:m=2}.  Mimas' longitude at epoch for each panel was obtained using the JPL (Jet Propulsion Laboratory) NAIF (Navigation and Ancillary Information Facility) kernel appropriate for the epoch represented by that panel.  In those fits, the parameters for the adopted $m=1$ and $m=3$ patterns from Table \ref{tbl:elem_m=1224} were accounted for and fixed, but no attempt was made to remove the large radial anomalies, as this part of the discussion concerns only the qualitative character of the variation.  In the figure, the amplitude variation is obvious and appears systematic, starting at about 50 km in mid-2005, decreasing to nearly zero in late 2006, then increasing back to about 70 km in early 2009.  Moreover, the orientation of the pattern relative to Mimas varies systematically, with Mimas initially lagging the pattern by $\sim$ 30$^\circ$, and eventually winding up leading the pattern by $\sim$ 10$^\circ$.

In our adopted solution, the variation in the amplitude and orientation of the total $m=2$ pattern seen in Fig. \ref{fig:m=2} arises from the interference of two $m=2$ patterns of comparable amplitude moving at slightly different speeds.  One mode is a response to the Mimas perturbation; the other is a free unstable mode like the $m=1$ and $m=3$ waves described in the previous two sections.  We held the speed of the forced $m=2$ pattern fixed at Mimas' average mean motion for the interval during which the observations were acquired.  According to the best available fit to Mimas' speed as a function of time (R.A. Jacobson, pers. comm.), which is dominated by its resonance with Tethys, its average mean motion should have been about 381.98401$^\circ$/day during that time.  However, the residuals to that fit indicate that Mimas' true speed was actually slower than that by about 0.0005$^\circ$/day.  The standard deviation in Mimas' speed variation during that interval is about ${10^{-4}}^\circ$/day.  Therefore, we took Mimas' mean speed to be 381.9835(1)$^\circ$/day.  For that pattern speed, the location of the Mimas 2:1 ILR is $a_M$ = 117555.8(1) km.  In the presence of collective effects and self-gravity, the half-width of the resonance zone is given by the first wavelength of the density wave generated by the resonance  (see Eq. \ref{eq:cp_a}).  For the range of densities computed for the outermost portion of the ring (Sec. \ref{sec:m=3}), that half-width is $\sim$ 60 -- 90 km.

The amplitude expected for the pattern forced by the Mimas resonance is $ae_M = |A/(a-a_M)|$, where \citep{Goldreich1982}:
\begin{equation}
	A = \frac{M_s}{M_p} \frac{\alpha a^2}{3 (m - 1)} %
	         \left.\left( 2m + \alpha\frac{d}{d\alpha} \right) %
	              b_{1/2}^{(m)}(\alpha)\right|_{a=a_M}. %
                                                                  \label{eq:AAA}
\end{equation}
Taking $\alpha = a/a_s$, where $a_s = 185539.5$ km is Mimas' orbital semimajor axis, and noting that the expression involving the Laplace coefficient $b_{1/2}^{(m)}(\alpha)$ evaluates to $\sim$ 2.38 (see Tables 8.1 and 8.5 of \citet{MurrayDermott}), the computed amplitude is $ae_M = $ 37.5 km, differing by about 8\% from the value of 34.6 km implied by the forced eccentricity given in Table \ref{tbl:elem_m=1224}.

The speed of the best-fit free $m=2$ pattern in Table \ref{tbl:elem_m=1224} corresponds to an ILR at an orbital radius of 117537.7(2) km, about 18 km interior to the Mimas resonance.  The beat period is $\pi/\Delta\Omega_p$ = 5.5(1) yr, where $\Delta\Omega_p$ = 0.0896(7)$^\circ$/day is the difference in pattern speeds between the two components.  The width of the cavity between the free $m=2$ ILR and the outer B ring edge is $117568(4) - 117537.7(2) = 30(5)$ km.  With $\Upsilon$ = 1/2, Eq. \ref{eq:cp_b} yields a surface mass density across the cavity of $\Sigma = 42(7)$ g cm$^{-2}$; for $\Upsilon$ = 1/4, we have $\Sigma$ = 84(7) g cm$^{-2}$.  Both values are nearly identical to those computed from the $m=1$ mode, and consistent with that for the $m=3$ mode.  Fig. S1 in the online supplement shows an animation of the ring edge with the radial locations of the free and forced ILRs and the mean edge overlain.

In their measurements of the B-ring edge taken from Cassini VIMS (Visual and Infrared Mapping Spectrometer) data sets, \citet{Hedman2010} noticed a similar variation in the $m=2$ pattern to that in Fig. \ref{fig:m=2}.  However, they interpreted the variation in the context of libration near a resonance instead of as the interference of two similar patterns as we have here.  They used the formalism for an isolated particle near resonance, as in \citet{Greenberg1973}, with resonant argument:
\begin{equation}
	\phi = \lambda - 2\lambda_s + \varpi, %
                                                                   \label{eq:a}
\end{equation}
where $\lambda$ and $\varpi$ are the mean longitude and periapse longitude of the particle respectively, and $\lambda_s$ is Mimas' mean longitude.  In that approach, the time-variable total eccentricity $e(t)$ is decomposed into free and forced components, $e_f$ and $e_0$ respectively.  The particle traces out a circle in a polar coordinate system with $e$ as the radial coordinate and the resonant argument $\phi$ as the angular coordinate, as shown in Fig. \ref{fig:hk}.  The phase angle $\theta$ increases at a constant rate $\dot\theta$, yielding the explicit time variation given by \citet{Hedman2010}):
\begin{eqnarray}
	e(t) & = & \sqrt{e_0^2 + e_f^2 + 2 e_0 e_f \cos\theta(t)} \nonumber \\ %
	\sin\phi(t) & = & \frac{e_f\sin\theta(t)}{e(t)} \nonumber \\ %
        \cos\phi(t) & = & \frac{e_f\cos\theta(t) + e_0}{e(t)} \nonumber \\ %
        \theta(t) & = & \theta(t_0) + \frac{\beta n}{e_0} t,  %
                                                                   \label{eq:fa}
\end{eqnarray}
where $\beta$ is the resonance strength.  The epoch $t_0$ is a constant of integration and the libration frequency $\dot\theta = \beta n/e_0$ depends on the forced eccentricity (i.e., it depends on the distance between the particle semimajor axis and the location of the exact resonance).  

To determine the explicit time variation in our interpretation, consider the fractional radial excursion for the composite $m=2$ pattern as a function of longitude $\lambda$ in a frame rotating with Mimas:
\begin{equation}
    \frac{\Delta r}{a} = e_0 \cos 2\lambda + e_f \cos 2[\lambda - \\  
               \Delta\Omega_p(t-t_0)], \\
							      \label{eq:bbb}
\end{equation}
where $e_0$ and $e_f$ are the eccentricities of the forced and free components respectively, and $\Delta\Omega_p$ is difference between the pattern speeds of the free and forced components.  The curve described by Eq. \ref{eq:bbb} is a quasi-sinusoid with frequency $2\Delta\Omega_p$ whose phase can be determined by finding an extremum, i.e., by solving $\partial(\Delta r/a)/\partial\lambda = 0$ for $\lambda$.  Identifying $2\Delta\Omega_p(t-t_0)$ with $\theta(t)$, and noting that $\lambda = \varpi$ at the minimum, we have:
\begin{equation}
    e_0\sin 2\varpi = -e_f\sin 2(\varpi - \theta) \\
      = -e_f\sin 2\varpi\cos\theta + e_f\cos 2\varpi\sin\theta, \\
							      \label{eq:ccc}
\end{equation}
or
\begin{equation}
    \tan 2\varpi = \frac{e_f\sin\theta}{e_0 + e_f\cos\theta}. \\
							      \label{eq:ddd}
\end{equation}
Moreover, from (\ref{eq:a}) we have:
\begin{equation}
	\lambda - \varpi = 2(\lambda - \lambda_s) - \phi. %
                                                                 \label{eq:eee}
\end{equation}
Since $\lambda_s$ can be taken to be zero in the rotating frame, and $\lambda = \varpi$ at periapse, we have $2\varpi - \phi = 0$, or $2\varpi = \phi$.  The total eccentricity $e(t)$ is simply the fractional amplitude (\ref{eq:bbb}) evaluated at $\lambda=\varpi$, and can be obtained by substituting (\ref{eq:ddd}) back into (\ref{eq:bbb}).  The time variation is therefore:
\begin{eqnarray}
     e(t) & = & \sqrt{e_0^2 + e_f^2 + 2 e_0 e_f \cos\theta(t)} \nonumber \\ %
     \sin \phi(t) & = & \frac{e_f\sin\theta(t)}{e(t)} \nonumber \\ %
     \cos \phi(t) & = & \frac{e_f\cos\theta(t) + e_0}{e(t)} \nonumber \\ %
     \theta(t) & = & 2\Delta\Omega_p(t-t_0).  %
                                                                \label{eq:faa}
\end{eqnarray}
Eqs. \ref{eq:faa} are identical to Eqs. \ref{eq:fa}, except that the frequency $2\Delta\Omega_p$ is independent of the distance between the radial locations of the streamline and the resonance, whereas in Eqs. \ref{eq:fa} $\dot\theta$ is uniquely determined by $e_0$, which is determined by that distance.  Indeed, our best-fit frequency of $2\Delta\Omega_p$ = 0.179(7)$^\circ$/day is not consistent with the computed frequency $\beta n/e_0$ = 0.128(1)$^\circ$/day.  Moreover, we have observed two other free unstable modes near the B-ring edge. Consequently, we believe that our interpretation of the observed time variation of the total $m=2$ pattern is more natural than that of \citet{Hedman2010}.  

\citet{Hedman2010} suggested that the motion of the $m=2$ pattern on the B-ring edge could be responsible for eccentric features found within various gaps in the Cassini Division.  In particular, it was noted that a model where the oscillation frequency of the $m=2$ pattern about the Saturn-Mimas line was $\sim$ 0.06$^\circ$/day was consistent with Cassini VIMS, RSS and historical occultation data and was also equal to the difference in pattern speeds between adjacent eccentric features in the Cassini Division.  Because that frequency is slower than our result by a factor of nearly 3, the libration model for the Cassini Division features may need to be modified for it to be successful.

At the central epoch of our fit, JED 2454256.9, Mimas' longitude was 10.1(1)$^\circ$, where the quoted uncertainty arises from the uncertainty in Mimas' mean speed, given above.  Comparison with Table \ref{tbl:elem_m=1224} shows that the forced $m=2$ pattern lagged Mimas by 2.9(3)$^\circ$.  That lag angle, $\Delta$, is diagnostic of the rate of angular momentum exchange between the ring and the satellite, allowing us to compute the torque exerted by the ring on Mimas \citep{Borderies1982}:
\begin{equation}
    T_s = (m a n)^2e\frac{M_s}{M_p}\Sigma a \Delta a |\sin m\Delta|, \\
							      \label{eq:Ts}
\end{equation}
and hence the rate of orbital evolution 
\begin{equation}
    \frac{da_s}{dt} \sim \frac{2 T_s}{n_s a_s}, \\
							      \label{eq:dasdt}
\end{equation}
assuming the torque acts only to change Mimas' semimajor axis and not its eccentricity.  Taking $\Delta a = ae$ and considering the range of mean surface densities 50--100 g cm$^{-2}$ computed above for the outermost zone, the resulting rate of orbital evolution $da_s/dt$ ranges from $\sim$ 0.06 to $\sim$ 0.12 cm yr$^{-1}$, amounting to about 3000--6000 km of orbital expansion if $T_s$ were to act continuously over the age of the solar system.  Balancing the satellite torque against the viscous torque delivering angular momentum to the ring edge, the kinematic viscosity $\nu$ is related to the lag angle by \citep{Borderies1982}:
\begin{equation}
    \sin m\Delta = \frac{\nu}{n(m a)^2}\left(\frac{M_p}{M_s}\right)^2, \\
							      \label{eq:nu}
\end{equation}
which yields a kinematic viscosity $\nu \simeq $ 20 cm$^2$ s$^{-1}$ for values appropriate for the B-ring outer edge.  This value for $\nu$ is on the low end of the various estimates obtained from spiral density waves in the A ring (see for example Fig. 20 of \citet{Tiscareno2007}).  Note, however, that this calculation assumes that the hydrodynamical approximation applies in the outer B ring.  In reality, a scalar parameter may be insufficient to represent the relation between viscous stress and deformation in this region\citep{Hahn2009}.

\section{Sites of massive bodies and vertical structure} 
Visual inspection of the thousands of Cassini images used in our kinematical analysis revealed distinct azimuthally limited regions in the resonance zone at the outer edge in which the ring appearance was visually disturbed and not azimuthally smooth as in other regions.  Two of these were most prominent.  The most obvious, which we have labeled region A (see Fig. \ref{fig:maps}), has been seen in several dramatic images taken as Saturn was approaching its August 11, 2009 northern vernal equinox in which vertical structures, the tallest 3.5 km, were clearly identifiable by their shadows (Figs. \ref{fig:thumbs}b and \ref{fig:thumbs}c).   Other images taken near equinox revealed the same region, spanning about 10$^\circ$ (or 20000 km) in azimuth, which was again betrayed by its shadows.  

In earlier low-sun conditions prior to the shadow-casting geometry, region A was captured in an ansa movie (Set 54) as a chevron-shaped structure passing around the ansa of the ring as seen from Cassini (Fig. \ref{fig:thumbs}f).  However, in yet earlier images taken under high sun, lit-face conditions, it wasn't as easily seen, explaining its late detection.  Instead, region A made itself apparent in high-sun conditions by a rise in brightness, usually in concert with a small discontinuity (or bump) in the variation of radius with azimuth, as seen in the maps shown in Fig. \ref{fig:maps}.

Fits to positional measurements of region A revealed that it is traveling at a uniform speed (Fig. \ref{fig:lonplots}a).  Assuming that the fitted motion corresponds to a simple, unperturbed Keplerian orbital speed, we can infer the region's precise orbital distance.  The fitted speed and inferred orbital distance of Region A were found to be 758.986(4) $^\circ$/day and 117547.6(5) km, about 8 km interior to the 2:1 resonance, with a J2000 longitude of 95(1)$^\circ$.  Region A is therefore situated within the 2:1 resonance zone, in which $\tau \sim$ 1 -- 1.2.   At these fairly high optical depths, any moderate differences in optical depths -- e.g., $\Delta\tau = 0.5$ -- are more easily distinguished when the observer is on the unlit side of the rings and the viewing angle is shallow; the effect is enhanced when the sun angle is also low.  This is a result of multiple scattering among the particles (and therefore diffuse transmission through the rings) becoming increasingly choked off with increasing optical path lengths; the contrast between regions having different optical depths is greatest at the high optical depths typical of the resonance zone of the outer B-ring edge.  (Compare, for example, Figs. 11 and 12 in \citet{Porco2008}).  Hence, even when the sun was not low enough to cause 4-km structures to cast shadows, region A was noticeable on the dark side of the rings, especially at low sun angles, by its unusual chevron-shaped variations in brightness.  However, in some images where region A should have been seen based on the illumination and viewing geometry and on its mean motion, it was not.   Apparently, its position with respect to the $m=2$ pattern is also a factor in is visibility: it appears more pronounced when it is near the steepest radial transition in the $m=2$ pattern (i.e., quadrature).

In all, 17 sightings of region A have been made in Cassini images spanning 3 years, from day 2006-247 to day 2009-233.  (In a few of these cases, the region's uniform motion was used to predict where it might be at an earlier time).

Another region, labeled B, was also noticed in the resonance zone (Fig. \ref{fig:maps}).   The degree to which the ring is disturbed in region B is not nearly as extreme as in region A.  It too was easier to see at low sun angles on the dark side for the same reasons as given above.  Sometimes, it too was seen as a slight, short-wavelength bump in the outer B-ring edge, and sometimes, when it should have been seen based on its motion and the illumination/viewing geometry, like region A it was not.  Over all, region B was sighted 10 times in images spanning 3.25 years, from day 2005-174 to day 2008-312.   Its uniform orbital speed, determined through least-squares fitting, was found to be 758.810(4) $^\circ$/day; its inferred orbital radius, assuming unperturbed simple Keplerian motion, is 117565.7(4) km, about 10 km exterior to the 2:1 resonance and very nearly coincident with the ring's outer edge (Fig. \ref{fig:lonplots}b).  Its J2000 longitude is 155(2)$^\circ$.

The fact that these locales are moving at Keplerian speeds over the course of several years, and are not traveling at the rotational speeds of any of the components in Table 1 implies that they are more likely to be the sites of more-massive-than-average bodies in the rings rather than densely compressed ``shocked'' regions created by the passage of ring material through the overall non-axisymmetric pattern.

We have rejected the possibility that the two observed regions are the sites of co-rotation resonances. First, there are no co-rotation resonances in the vicinity of the outer edge of the B ring with known moons big enough to have a pronounced effect.  It would be more likely that the populations are in resonance with each other, perhaps similar to the behavior of Janus and Epimetheus, or perhaps even similar to Jupiter and its Trojan asteroids.  In the latter case, the less massive population would be situated at either the L4 or L5 position of the more massive population; in the former, it would be librating in a horseshoe orbit around the L3 position.  However, a look at how the difference in their longitudinal positions varies with time (Fig. \ref{fig:lonplots}c) (which requires short-range extrapolation of the position of one or the other region since at any given instant, only one region was observed) shows that libration is not likely.   It is apparent from Fig. \ref{fig:lonplots}c that the two regions coincided in azimuth in mid-2006.  While no observations of region A exist prior to this time, observations of region B do.  If region B were librating, its mean motion would have changed across the conjunction.  Instead, its motion from 2005 through 2006 is consistent with uniformly increasing longitude with increasing time.  

Therefore, the simplest interpretation is that these two regions are locales where one or more bodies are located, with each group orbiting independently of the other on two distinctly separate orbits.  These bodies are apparently massive enough to alter the ring material around them considerably, though region A's effect is clearly far larger. 

Despite the fact that the Cassini equinox imaging captured region B, the only place around the B-ring edge where vertical structures have been seen casting shadows is region A.  The shadow orientations here indicate the vertical structures are likely coincident with the bright regions and streaks lining the B-ring's outer edge (Fig. \ref{fig:thumbs}c).  This vertical structure is likely due to "splashing" \citep{Borderies1985} as particles are forced over one another to accommodate severe radial compression of streamlines in the already closely-packed ring.   

This effect may very well be far more pronounced around large bodies in the rings, where the large size of the body causes streamlines to divert and compress around it even more.  Bodies hundreds of meters in size in the A-ring can create three-dimensional propeller-type features in which surrounding ring particles are perturbed hundreds of meters in the vertical direction \citep{Tiscareno2010}.  It is plausible that, at the outer B-ring edge where the ring's density is high and resonant forcing is extreme, vertical motions surrounding bodies hundreds of meters in size can be even greater, and the extreme heights of the observed vertical structures may indicate the presence of one or more bodies of kilometer-class size.  The additional possibility, that the body (or bodies) in region A are on inclined orbits and are driving vertical motions the way that Daphnis drives vertical motions in the edge of the Keeler gap \citep{Weiss2009}, is not likely since sustained vertical motions, even for a 1-km moonlet, would almost certainly damp due to the collisions with dense ring material confined to the ringplane.

Thus, we speculate that at the heart of region A is a large body or bodies, hundreds of meters to perhaps kilometers in size, and massive enough to severely disturb and elevate to significant heights the material passing around it.  Region B is likely also the site of larger than average body or bodies, but not large enough to force large vertical motions despite its presence in the strongest resonance zone (and therefore most extreme compressive environment) in Saturn's rings.

\section{Moonlet S/2009 S1} 
A new object, provisionally labeled S/2009 S1, was discovered \citep{Porco2009} in image N1627301569, taken through a clear filter with an exposure of 820 ms with the spacecraft and Sun on the same side of the rings (Fig. \ref{fig:thumbs}g).  Based on the length of the shadow -- $\sim$ 36 km -- and
the width of the bright feature -- $>$ 1 km  -- it is possible that we are
seeing the innermost core of a propeller-shaped disturbance created by an
embedded moonlet \citep{Tiscareno2006}, with a height above the ringplane of
$\sim$ 150 meters and therefore a diameter of $\sim$ 300 m.  We determined its planetocentric distance (and presumably its orbital radius) to be 116914(17) km, about 650 km interior to the ring edge.  No standard fiducials were available in the discovery image, so the orbital radius was estimated by scaling to two nearby features whose radii were linearly extrapolated from earlier frames in the sequence in which circular reference features were present.  The uncertainty in that estimate arises primarily from the uncertainty in the radii of the fiducial features.  

The interpretation of S/2009 S1 as a solid body, surrounded by the core
of a propeller disturbance, is supported by the presence of a shadow.  If the observed feature were, say, an impact plume observed very shortly after its creation, so that differential motion had not yet sheared it out as has been observed for other impact clouds imaged with Cassini (http://www.ciclops.org/view.php?id=5782), it could perhaps be optically thick enough to cast a shadow.  However, no shadows have been observed accompanying any known impact plumes observed by Cassini.

S/2009 S1 is almost certainly long-lived.  An object 300 m across could not have formed recently from accretion in the outer B ring whose present-day vertical thickness is $\sim$ 5 m \citep{Colwell2007}.  And even in the past, when the ring system was likely much thicker, a solid, denser core nearly the size of the object itself would have been required to initiate accretion, especially this close to Saturn (See Fig. 5 in \citet{Porco2007}).  Hence, we think it is safe to conclude that S/2009 S1 is predominantly a solid moonlet, perhaps with a thin rubble outer shell, that likely dates back to the origin of the ring system.

The presence of this moonlet lends support to the idea put forth in the previous section that embedded moonlets exist near the edge of the B ring.  In a disk of uniform surface density, objects that are too small to open a gap typically drift inwards as the torque from material inward of the body is slightly stronger than that due to material exterior to the body (i.e, type-I migration).  In the presence of a negative surface density gradient, $d\Sigma/dr$, the net torque can be reversed resulting in outward migration \citep{Ward1986}.  It is therefore plausible that objects of this size, and perhaps larger, migrated across the region of the outer B ring early in the history of the rings to become trapped permanently in the Mimas 2:1 resonance.  

The alternative hypothesis, where the objects accreted from local ring material within the resonance zone \citep{Esposito2009}, seems unlikely because the dynamical stirring in this region is more than sufficient to overcome the self-gravity of the ring particles.  The escape velocity for a spherical particle is:
\begin{equation}
    v_\mathrm{esc} = \left(\frac{G M}{R}\right)^\frac{1}{2} = \left(\frac{8}{3}\pi\rho G\right)^\frac{1}{2} R, \\
							      \label{eq:v_esc}
\end{equation}
where $M$, $R$, and $\rho$ are the mass, radius and density of the particle respectively.  For a 10-m particle composed of solid ice, that yields $v_{esc} \lesssim 1$ cm s$^{-1}$.  The velocity dispersion in the resonance zone likely arises primarily from the large-amplitude radial oscillation of the ring edge: 
\begin{equation}
    \frac{\partial{r}}{\partial{t}} \sim \frac{\Delta r}{\Delta t} = \frac{n m}{2\pi}\Delta r,\\
							      \label{eq:drdt}
\end{equation}
where $\Delta r$ is the peak-to-peak variation of the radial oscillation.  For $\Delta r \sim$ 100 km, the radial velocity is $\sim$ 5 m s$^{-1}$.  Even with a coefficient of restitution as low as 0.02 (Figure 22 of \citet{Porco2008}), the resulting velocity dispersion is $\sim$ 10 cm s$^{-1}$, much faster than the escape velocity of the largest particles.  Therefore, as there's no reason to expect particles in this region to be especially sticky, we see no reason to argue for accretion in this region as opposed to other less perturbed regions in the rings.  The fact that we find massive bodies within the strongest resonance zone in Saturn's rings is more plausibly explained as a result of migration and resonance trapping, analogous to the process, prevalent throughout the solar system, by which satellites and even planets have migrated into, and become trapped, in resonances \citep{Lee2009, Ferraz-Mello2003}. 

The new moonlet lacks the propeller-shaped disturbance, with extensive wings, that is associated with embedded moonlets in the A ring \citep{Tiscareno2006}.  The absence of extensive wings may not be surprising because the ring material in a dense ring like the B ring would be expected to fill in any gaps around the moonlet more quickly than in a less dense region like the mid-A ring.  Alternatively, it may simply be that the propeller feature is present, but does not comprise a strong enough optical depth contrast to be observable in this optically thick part of the rings, or that it is not resolved.

\section{Relation between embedded moonlets and kinematics} 
The presence of massive bodies orbiting just interior to the B-ring outer edge may explain the largest discrepancies between the four-component kinematical model and the measurements in Fig. \ref{fig:plots}.  The deviations from the kinematical model (most obvious in the $m=3$ plot) are dominated by relatively narrow radial excursions with azimuthal widths around 20--30$^\circ$.  Comparing the longitudes of these ``spikes'' with those of the disturbed regions, it was apparent that many of them coincide with the measured locations of regions A and B.  Indeed, the association is unambiguous in some cases (e.g., sets 2, 3, 54, 64; see Fig. \ref{fig:maps}\footnote{Note however, that set 64 is not plotted in Fig. \ref{fig:plots} for reasons stated earlier}), as those features either were initially identified based on their anomalous radial variation, or an anomalous radial excursion is an obvious part of the observed feature.  Moreover, most of the spikes that are not coincident with the region-A and -B sightings are coincident with their {\it computed} locations based on the the orbits derived from those measurements.  Therefore the kinematical modeling allows us to identify additional occurrences of those disturbances that we were not able to identify from inspection of the images.  In Fig. \ref{fig:plots}, the spikes that are coincident with one or the other region are colored red (region A) or blue (region B) in the $m=3$ plot.  In total, 13 radial spikes were associated with region A, and 5 with region B.  Three additional prominent spikes that coincide with neither region are colored green in the figure, and may represent one or more additional sites.  Indeed, the original inspection and analysis of the images that revealed regions A and B turned up a few other candidates, but none that could be linked together with a common mean motion in the way that regions A and B could.  Other less prominent features may also be identified as spikes, or they may be unrelated, as the residuals contain quite a bit of structure at the 10-km level.  Because the spikes apparently arise from a different process than that driving the normal mode patterns, those measurements were excluded from the adopted fit in Table \ref{tbl:elem_m=1224}.  

From the elements in Table \ref{tbl:elem_m=1224}, we calculate that the variation in the amplitude of the observed $m=2$ pattern reached a minimum value around mid-2006.  It is an interesting coincidence that regions A and B would have had a conjunction at about the same time, as seen in Fig. \ref{fig:lonplots}c.  Moreover, as can be seen in Fig. \ref{fig:plots}, most of the large spikes appear to align with the $m=3$ pattern, even though we would expect no correlation between the spikes and the phase and speed of the pattern in the fitting process because those spikes were not considered in the kinematical fit.  Additional work will be required to understand the significance, if any, of these coincidences, but they seem to imply a complicated set of commensurabilities between the wave patterns at the edge of the ring and the motions of the embedded mass concentrations.

\section{Conclusions} 
We have confirmed conclusions based on Voyager results, and Cassini occultation results, that the outer edge of the B-ring is controlled only in part by the Mimas 2:1 ILR.  We have extended and improved earlier work using 4 years of Cassini images to show that, as others recently have found \citep{Hedman2010}, the $m=2$ distortion of the edge is undergoing large variations in amplitude and in orientation with respect to the direction to Mimas.  However, in addition to the forced $m=2$ component and unlike previous interpretations, our results are consistent with the presence of at least three prominent unstable {\it free} normal modes with azimuthal wavenumbers $m=1$, $m=2$, and $m=3$.  The forced $m=2$ component is rotating with Mimas' mean motion; one of its periapses is fixed relative to the satellite, lagging by $\sim 3^\circ$, and a simple linear superposition of the free and forced $m=2$ components creates the variability in the $m=2$ shape of the ring.  

The measured $\sim$ 3$^\circ$ lag angle implies that the torque exerted on Mimas by the ring is sufficient to expand its orbit by as much as $\sim$ 0.12 cm yr$^{-1}$.  With that lag angle as a constraint, forward modeling using a model along the lines of \citet{Hahn2009} could allow for a precise estimate of the ring's viscosity; a rough estimate gives $\nu \simeq $ 20 cm$^2$ s$^{-1}$.

We have shown that the $m=1$ normal mode, with a radial amplitude of $\sim 21$ km, is likely a free standing density wave trapped in a resonant cavity at the outer edge of the B-ring.  This is the first observation of a long-wavelength ($\Lambda \sim$ 505 km) unforced wave in Saturn's rings, and points to significant amplification of the $m=1$ wave as it reflects off the B ring's sharp outer edge.  The $m=2$ and $m=3$ disturbances have smaller wavelengths of $\Lambda \sim 60$ km and $\Lambda \sim 48$ km, respectively, and their amplitudes are $ae \sim 37$ km and $ae \sim 12$ km.  An $m=1$ free mode at the outer edge of the B ring was predicted by \citet{Borderies1985} based on the availability of a sharp edge and the inference of viscously overstable conditions in this high surface density region.  The discovery of the $m=1$ and the other free modes supports this interpretation and strongly suggests that viscous overstability is responsible for most, if not all, of the unforced, structures throughout the B ring and other dense ring regions, like the inner A ring, over a broad range of spatial scales.  Moreover, our results suggest that brightness variations in the outer $\sim$ 250 km of the B ring may be primarily due to trapped free density waves rather than albedo variations, as suggested by \citet{Cuzzi1984}.
  
Because the B-ring edge can be modeled using a simple linear superposition of modes, and because the observed value for the forced $m=2$ amplitude is nearly identical to its predicted value, we argue that any nonlinear coupling between the observed modes is weak, though interaction between the two $m=2$ modes may be enhanced during the relatively long interval when the two patterns are nearly aligned.  At that time, when the total $m=2$ amplitude is the greatest, the nonlinearity parameter $q = a (de/da)$ is also at its greatest and may lead to saturation of the free mode, limiting its growth.  

The free modes we have found here are all nodeless: i.e., the sign of the radial displacement is constant across the resonant cavity.  Indeed, different modes with the same $m$ but with radial nodes within the resonant cavity and hence with shorter wavelengths are possible and in principle could be present.   However, for the same radial displacements, these shorter-wavelength modes would have higher nonlinearity parameters $q$ than the nodeless modes so their interactions with the nodeless modes of the same $m$ would therefore be expected to be nonlinear and lead to damping; this would be especially so for the $m=2$ and $m=3$ modes.  Consequently, it is plausible that the smaller amplitude modes with radial nodes, especially the $m>1$ modes, have been significantly suppressed by their more robust nodeless counterpart.  

Nonetheless, some of these modes, if present, would have pattern speeds close enough to those measured here that they would not be detectable without a longer time baseline of observations.  Additional data sets will be obtained at regular intervals throughout the Cassini Solstice Mission:  during the next 7 years we hope to better than double the time baseline of this investigation.  Also, by using the orbit fits to regions A and B, we may now hope to unravel the historical occultation data sets dating back to the Voyager era, which have been difficult to model in the past because the smoothly varying normal-mode pattern was interrupted in a heretofore unpredictable way by the spikes associated with these regions.

The narrow, simply eccentric ringlets seen throughout Saturn's rings and the rings of Uranus are also likely the results of viscously unstable density waves of $m=1$ reflecting off the sharp edges of a resonance cavity \citep{Borderies1985}. The $\sim 60$-km amplitude $m=1$ narrow Maxwell ringlet in Saturn's C ring is an example.  However,  the existence of lower- and higher-wavenumber modes are also possible within a narrow ring: the Huygens ringlet, a roughly 50-km-wide ringlet a few hundred km exterior to the B-ring edge, possesses an $m=2$ mode with an amplitude of $\sim$ 20 km \citep{Porco1990}, and possibly a 1-km $m=6$ mode \citep{Spitale2006a}, on top of a 30-km amplitude $m=1$ mode, and some of the rings of Uranus exhibit $m=0$ and $m=2$ modes \citep{Porco1990}.  The detailed analysis of the Huygens and Maxwell ringlets will be a topic of future work.  

The three free modes discovered in this work reside in zones ranging in width from 24 km to 250 km, with outer boundaries at the B-ring edge.  Independent estimates for the average surface mass densities over those respective regions, for a given fixed value of $\Upsilon$, give consistent values.  We take this as support for the existence of these free modes, though if $\Upsilon$ were strongly dependent on wavenumber, that argument would be invalid.  A lower bound on the surface density was obtained by setting $\Upsilon=1/2$, yielding a weighted mean of $\sim$ 44 g cm$^{-2}$.  The true density could be near 100 g cm$^{-2}$, or perhaps even higher in the outermost zone.

Normal modes of the kind we have found at the edge of the B-ring are analogous to the free unstable modes that arise in numerical simulations of other celestial disks, such as protoplanetary disks \citep{Laughlin1997} and spiral galaxies.  However, definitive identification of such free modes in real spiral disk galaxies, and their role in the creation of grand design spirals, is hindered by the enormously long periods anticipated for such modes, on the order of $10^8$ yr, which cannot be directly measured.  Our results underscore the natural connection between Saturn's rings, our solar system's most accessible celestial disk, and other disks throughout the cosmos -- galaxies as well as protoplanetary disks.

We have found circumstantial evidence for the presence of large bodies in the resonance zone at the outer edge of the B-ring that are observable through their disturbance to the surrounding ring.  These disturbances are following simple uniform motion and are in orbits separated by 18 km.  They are responsible for most of the prominent short-wavelength fluctuations in radius that are observed at the ring edge; other regions containing larger-than-average bodies are also likely present and may be responsible for other short-wavelength disturbances seen in the outer edge.  The suggestion that moonlets are embedded near the outer edge of the ring is supported by the discovery of a 300-m diameter moonlet orbiting about 650 km interior to the ring edge.  Such bodies may have been present in the vicinity of the outer B-ring edge early in the history of Saturn's rings, subsequently migrating into the resonance zone to become permanently trapped in the Mimas 2:1 resonance.  This result suggests that radial spikes noticed in our recent work on the A-ring outer edge \citep{Spitale2009} could point to the presence of embedded moonlets there as well, likely trapped in the Janus/Epimetheus 7:6 inner Lindblad resonance that controls and shapes that edge.  This will be a subject of future work.


\section{Acknowledgements} 

We acknowledge the staff members within CICLOPS for their dedication and efforts in the design, execution, and processing of the ring imaging sequences used in this work.  We are indebted to P. Goldreich for valuable suggestions in the kinematic modeling and interpretation of our results.  We also thank R. A. Jacobson for sharing his fits to Mimas' mean motion, as well as R. Greenberg and J. Hahn for helpful discussions.  JNS and CCP were both supported by NASA and the Cassini project.  Finally, we thank M. S. Tiscareno for a thorough review that improved the quality of the text significantly.

\pagebreak


\clearpage

\section*{TABLES}
\begin{table}
 \begin{tabular}{llllll}
Set & Date & $\Theta_0$ ($^\circ$) & $\Theta$ ($^\circ$) & $g$ ($^\circ$) & Scale (km pixel$^{-1}$) \\ \hline

2$^a$ & 2005-174 & 111 & 71--72 & 45--48 & 9.8--10.7 \\
3$^b$ & 2005-176 & 111 & 65--70 & 20--29 & 5.7--21.3 \\
5$^b$ & 2005-231 & 111 & 64--73 & 3--13 & 3.7--21.7 \\
6$^a$ & 2006-247 & 106 & 85--85 & 163--164 & 12.5--13.1 \\
8$^a$ & 2006-312 & 105 & 24--45 & 34--103 & 2.1--3.0 \\
9$^a$ & 2006-322 & 105 & 84--88 & 143--148 & 6.5--7.2 \\
23$^a$ & 2007-099 & 103 & 54--68 & 18--23 & 3.9--4.9 \\
26$^a$ & 2007-118 & 103 & 80--81 & 46--47 & 9.7--10.0 \\
43$^b$ & 2008-028 & 99 & 30--68 & 25--84 & 1.8--4.6 \\
46$^a$ & 2008-097 & 98 & 65--67 & 31--32 & 8.8--21.2 \\
33$^b$ & 2008-151 & 97 & 63--69 & 31--35 & 6.9--30.5 \\
34$^b$ & 2008-159 & 97 & 52--56 & 41--44 & 5.6--14.6 \\
35$^b$ & 2008-215 & 96 & 53--61 & 36--44 & 6.1--22.2 \\
50$^b$ & 2008-237 & 95 & 51--59 & 37--46 & 5.9--17.8 \\
36$^b$ & 2008-312 & 94 & 38--48 & 47--56 & 5.6--10.7 \\
37$^b$ & 2008-343 & 94 & 20--36 & 58--74 & 4.1--6.3 \\
53$^a$ & 2009-044 & 93 & 64--83 & 151--162 & 4.9--4.9 \\
54$^a$ & 2009-056 & 93 & 75--88 & 159--162 & 4.9--4.9 \\
56$^c$ & 2009-098 & 92 & 38--39 & 84--84 & 6.9--7.1 \\
64$^c$ & 2009-207 & 90 & 53--64 & 140--151 & 1.8--6.4 \\
67$^c$ & 2009-228 & 90 & 77--77 & 111--111 & 12.6--12.6 \\
68$^c$ & 2009-231 & 90 & 79--79 & 119--119 & 13.6--19.2 \\
69$^c$ & 2009-232 & 90 & 80--80 & 122--122 & 13.5--13.5 \\
71$^c$ & 2009-233 & 90 & 82--82 & 127--127 & 12.9--12.9 \\ \hline
 \end{tabular}

\caption{Details of data sets used in this study.  The incidence, emission, and phase angles are represented by $\Theta_0$, $\Theta$, and $g$, respectively.  $\Theta_0$ and $\Theta$ are referenced to the northern ringplane normal.  Notes: (a) Ansa movie.  (b) Azimuthal scan.  (c) Used for shadows only.\label{tbl:data}}

\end{table}

\begin{table}

 \begin{tabular}{llllllll}
$m$ & $e \times 10^{-4}$ & $\varpi^{(m)}_0$($^\circ$) & $\Omega_p$($^\circ$/day) & $a_l$(km) & $\Upsilon\Lambda$(km) & $\Sigma_{\Upsilon=1/2}$(g cm$^-2$) & $\Sigma_{\Upsilon=1/4}$(g cm$^-2$) \\ \hline

 1	& 1.78(3)	& 73(1) 	& 5.098(3)	& 117314.92(8)	& 253(4)		& 43.5(7)	& 87.0(6) \\
 2	& 2.94(3)	& 7.2(3)	& [381.9835]	& 117555.8(1)	& --			& --		& -- \\
 2	& 3.17(3)	& 123.1(3)	& 382.0731(7)	& 117537.7(2)	& 30(5)			& 42(7)		& 84(7) \\
 3	& 1.00(2)	& 23.8(4) 	& 507.700(1)	& 117544.2(4)	& 24(5)			& 51(11)	& 100(11) \\

 \end{tabular}

\caption{Adopted best-fit streamline elements for a four-component model of the B-ring edge, along with inferred geometry and surface densities.  The semimajor axis is $a_0=$117568(4) km, where the quoted uncertainty includes the 0.7-km uncertainty in the radial scale.  Also, $\chi^2$/DOF = 63 and RMS = 7.9 km.  Longitudes are given at the central epoch, JED 2454256.9.  Quantities in brackets were held fixed.  Resonant locations $a_l$ were computed from the fit pattern speeds $\Omega_p$.  $\Upsilon\Lambda$ is the cavity width $a_0 - a_l$.  Surface density estimates are given for he assumption of a fixed edge ($\Upsilon=1/2$), and for a likely value for a free edge ($\Upsilon = 1/4$). \label{tbl:elem_m=1224}}

\end{table}

\pagebreak

\section*{FIGURE CAPTIONS}
\figcaption{Map projected mosaics (radius vs. longitude) for the Cassini data sets -- either azimuthal scans or ansa-staring movies -- used in this study.  The first image in each data set is assigned a longitude of zero.  Later images are mapped successively to the left or right (i.e., decreasing or increasing in co-rotating longitude from 360$^\circ$) according to their inertial longitudes, corrected for pattern rotation at Mimas' average mean motion of 381.9835$^\circ$/day, and the direction in which the scan was performed. Particles move through the mosaic from left to right.  Data set numbers appear to the right of each mosaic.  For each map, radii range from 117425 to 117690 km.  Locations of regions A and B (see text) are marked; labels in parentheses indicate the predicted location of the feature, though it was not observed. \label{fig:maps}}

\figcaption{Radial excursion vs. true anomaly for each component of the four-component model fit given in Table \ref{tbl:elem_m=1224}.  The true anomaly is $\lambda - \varpi^{(m)}$, where $\lambda$ is the measured longitude of the data point and the instantaneous orientation of the pattern is $\varpi^{(m)} = \varpi^{(m)}_0 + \Omega_p t$.  The periapse longitude at epoch, $\varpi^{(m)}_0$, and the pattern speed, $\Omega_p$, are independent parameters in the fit (whether fixed or free).  In each panel, only the residuals from the panel above are plotted.  In the $m=3$ plot, deviations described as ``spikes'' in the text are indicated by coloring those data points red for region A, blue for region B, or green for neither region.\label{fig:plots}}

\figcaption{Radial excursion vs. true anomaly for total $m=2$ fits to each data set used in this study.  The $m=1$ and $m=3$ patterns from Table 2 are present, but not shown.  The solid line shows the best-fit model with $a$, $e$, and $\varpi^{(m)}_0$ free; $\Omega_p$ was fixed at Mimas' average mean motion.  Dashed vertical lines indicate the position of Mimas (M) at each epoch.  Note that fixing $\Omega_p$ at Mimas' speed does not imply that Mimas should maintain a constant phase in these plots because $\varpi^{(m)}_0$ (i.e., the orientation at epoch) is a free parameter in each fit.\label{fig:m=2}}

\figcaption{Expected behavior of $e$ and $\phi$ for various combinations of free ($e_f$) and forced ($e_0$) eccentricities.  (a) Libration occurs when the trajectory does not encompass the origin, i.e., $|e_0| > e_f$.  Interior to the resonance ($e_0 > 0$) the particle traces a counter-clockwise circle in polar coordinates, thus librating about $\phi=0$.  Exterior to the resonance ($e_0 < 0$), the situation is reversed.  (b) Circulation occurs for $|e_0| < e_f$.\label{fig:hk}}

\figcaption{a--e) Region A at the outer edge of the B-ring in various ISS images.  f) ISS image of region B.  g) Moonlet S/2009 S1. \label{fig:thumbs}}

\figcaption{ a) Longitude vs. time for region A.  Solid line is the computed orbit, black diamonds are measured region locations with accompanying set numbers, colored lines with bars are measured spikes that correlate with region A.  Numbered spikes are relative to computed region locations because the region location was not measured for that data set.  Times are wrapped into a single period.  b) Same as (a) for region B.  c) Longitude difference between regions A and B vs. time.  Because regions A and B were never measured at precisely the same time, each point was computed using one measured and one modeled longitude: diamonds indicate that longitude A was measured; squares indicate that region B was measured.\label{fig:lonplots}}

\begin{figure}[h!]
 \begin{center}
{\scalebox{0.7}{\includegraphics{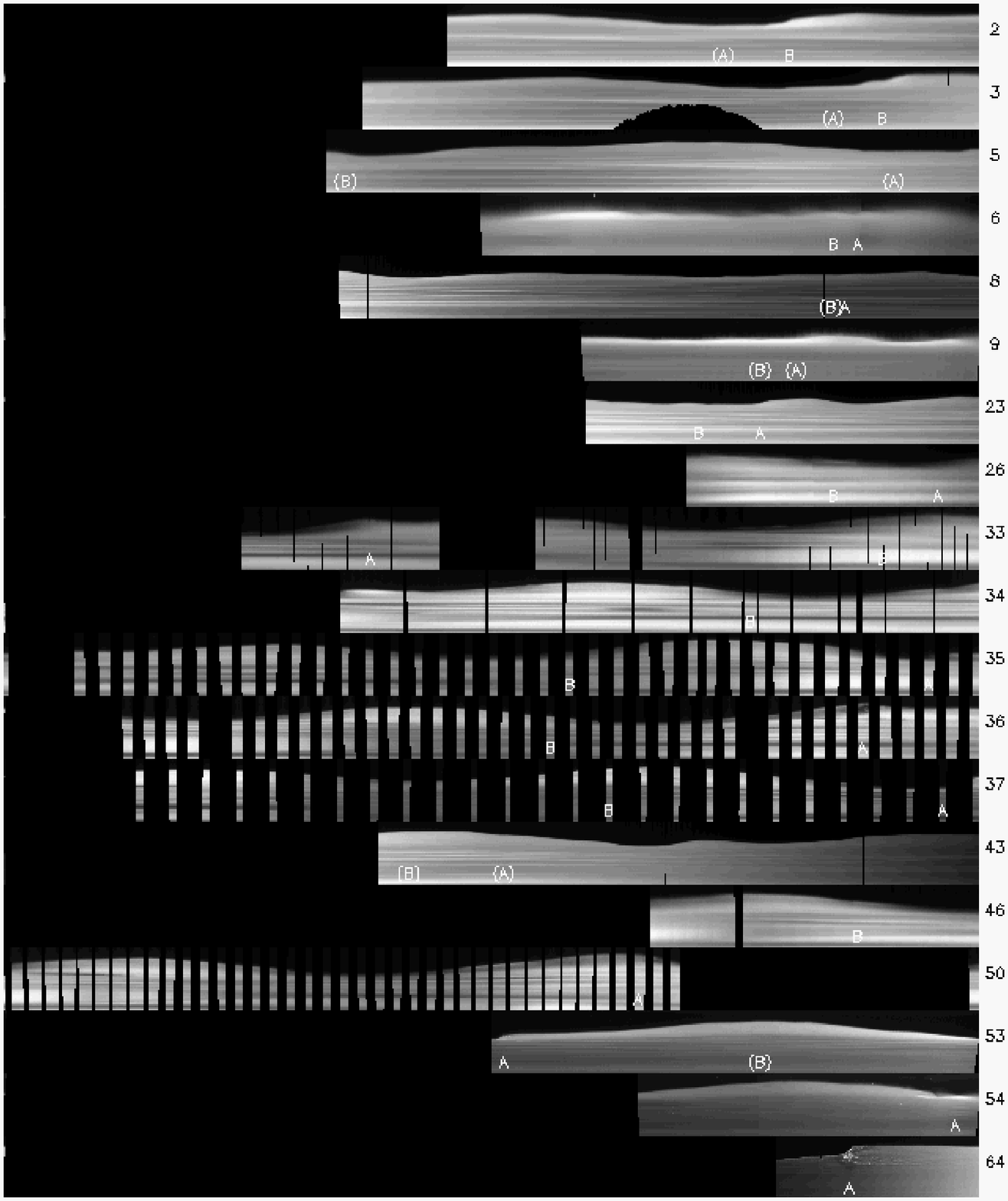}}}
 Fig. \ref{fig:maps}
\end{center}
\end{figure}

\begin{figure}[h!]
 \begin{center}
  {\scalebox{1.0}{\includegraphics{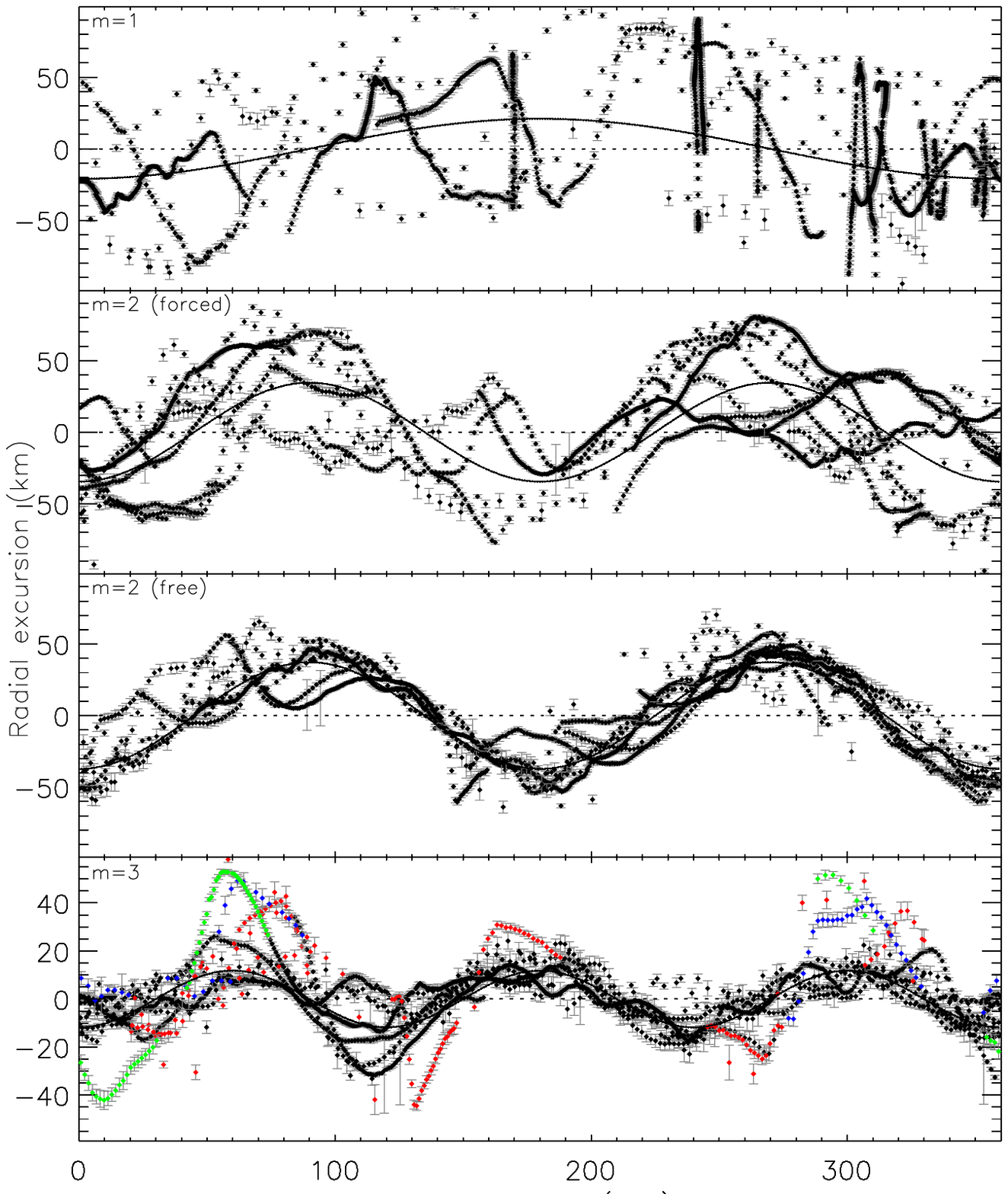}}}
 Fig. \ref{fig:plots}
 \end{center}
\end{figure}

\begin{figure}[h!]
 \begin{center}
  {\scalebox{0.89}{\includegraphics{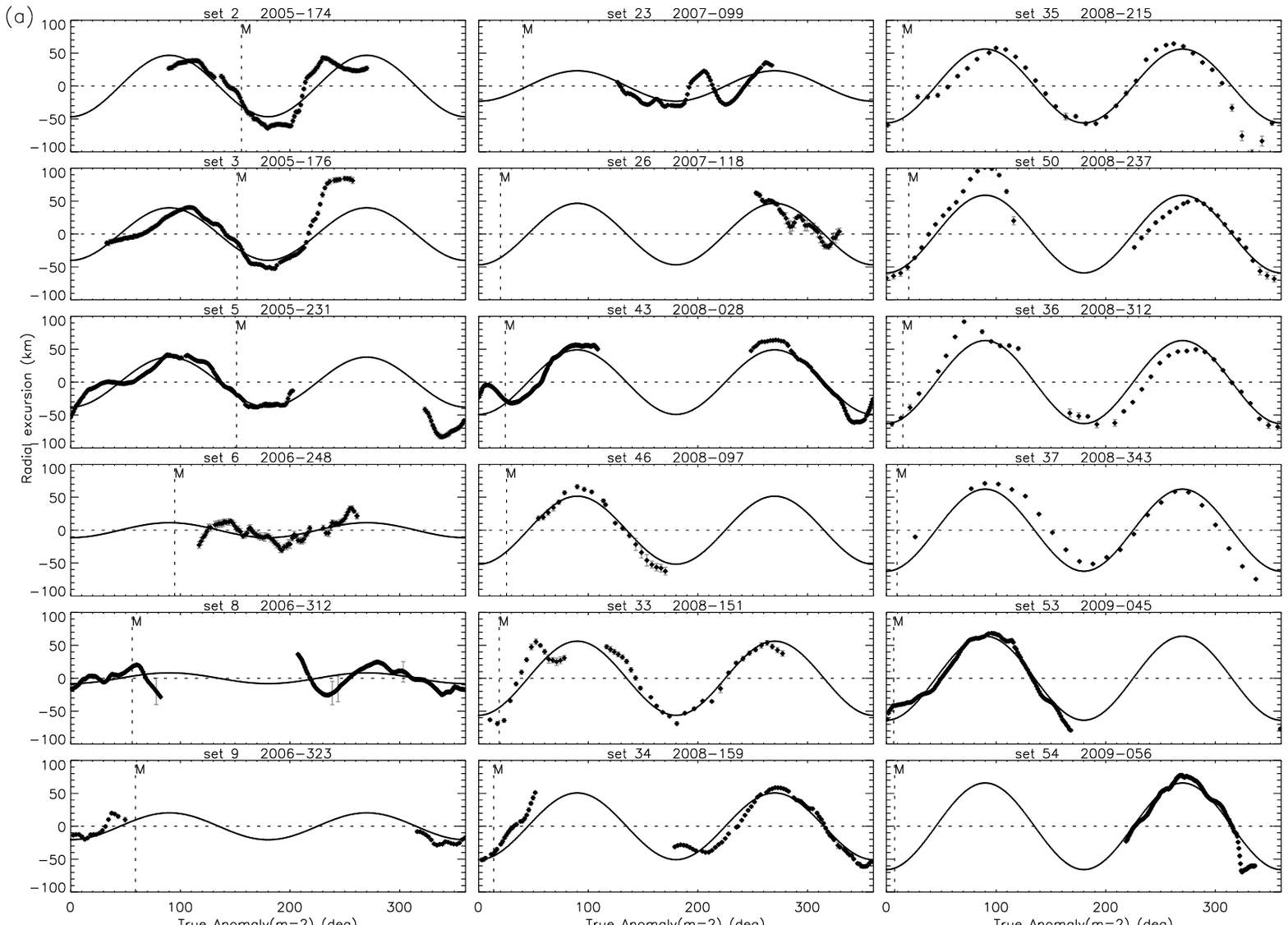}}}
 Fig. \ref{fig:m=2}
 \end{center}
 \end{figure}

\begin{figure}[h!]
  {\scalebox{0.7}{\includegraphics{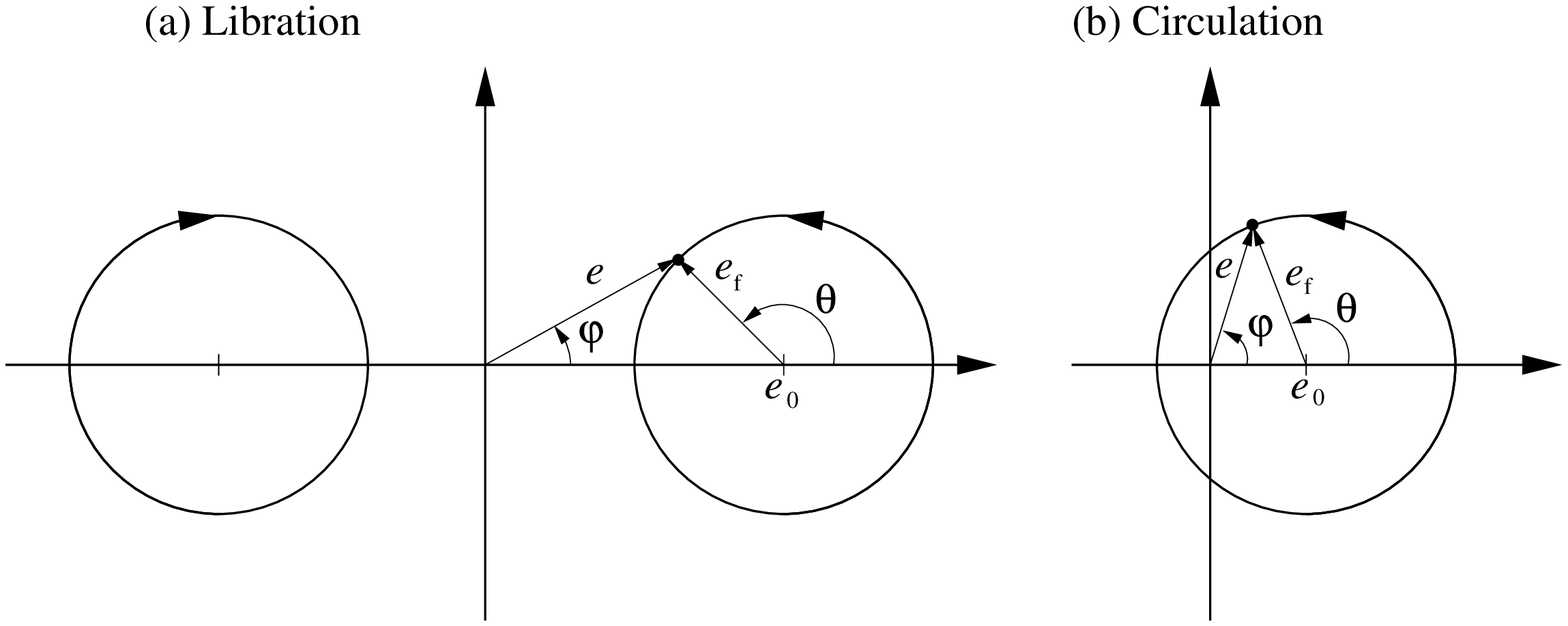}}}
 \begin{center}
 Fig. \ref{fig:hk}
 \end{center}
\end{figure}

\begin{figure}[h!]
 \begin{center}
  {\scalebox{0.6}{\includegraphics{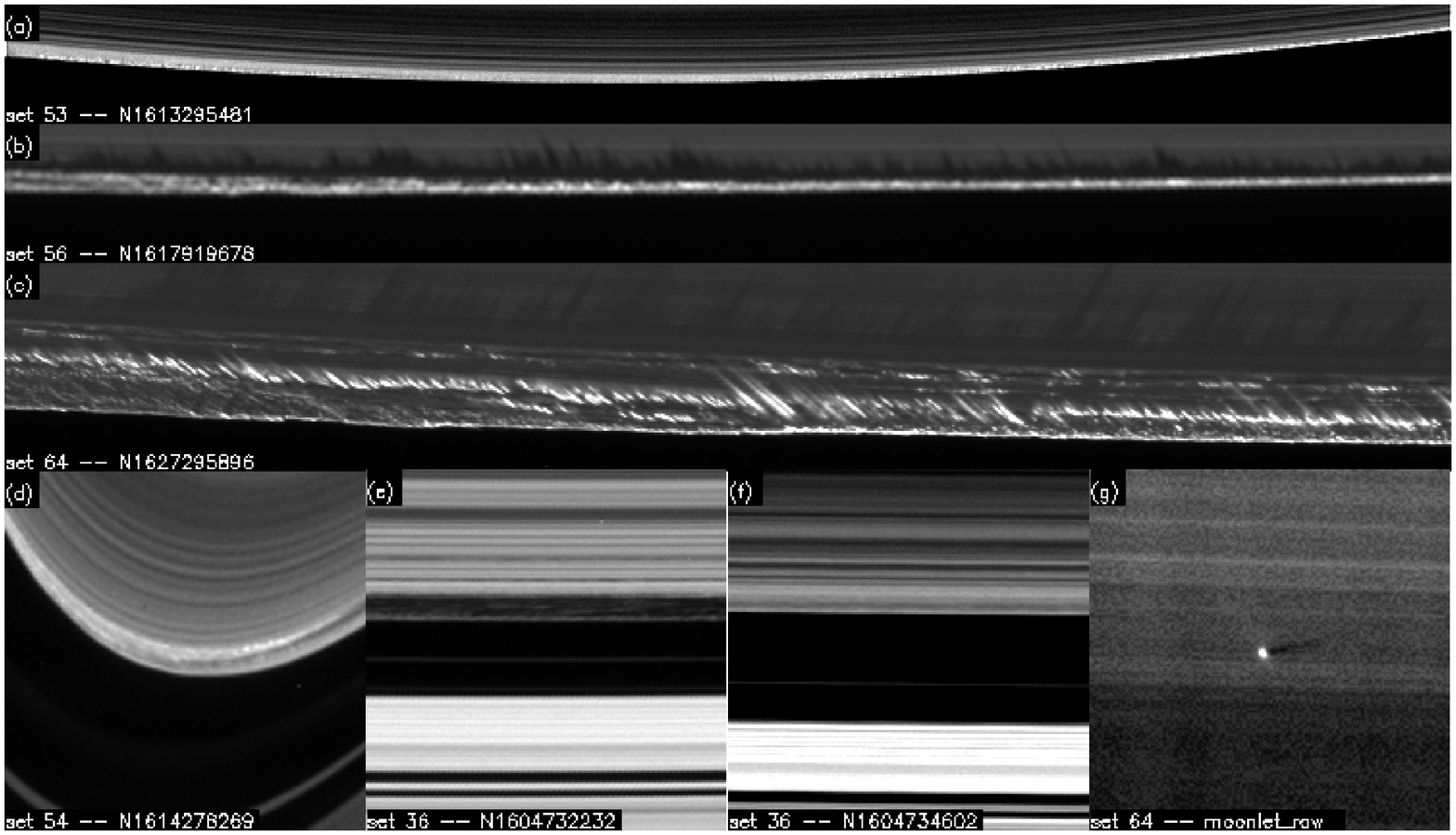}}}
 Fig. \ref{fig:thumbs}
 \end{center}
\end{figure}

\begin{figure}[h!]
 \begin{center}
{\scalebox{0.9}{\includegraphics{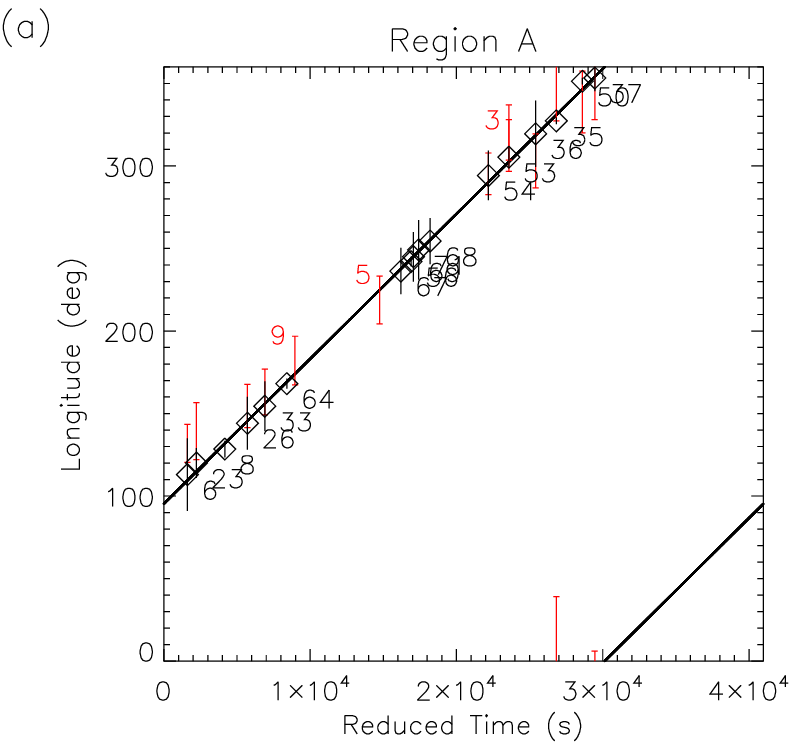}}}
{\scalebox{0.9}{\includegraphics{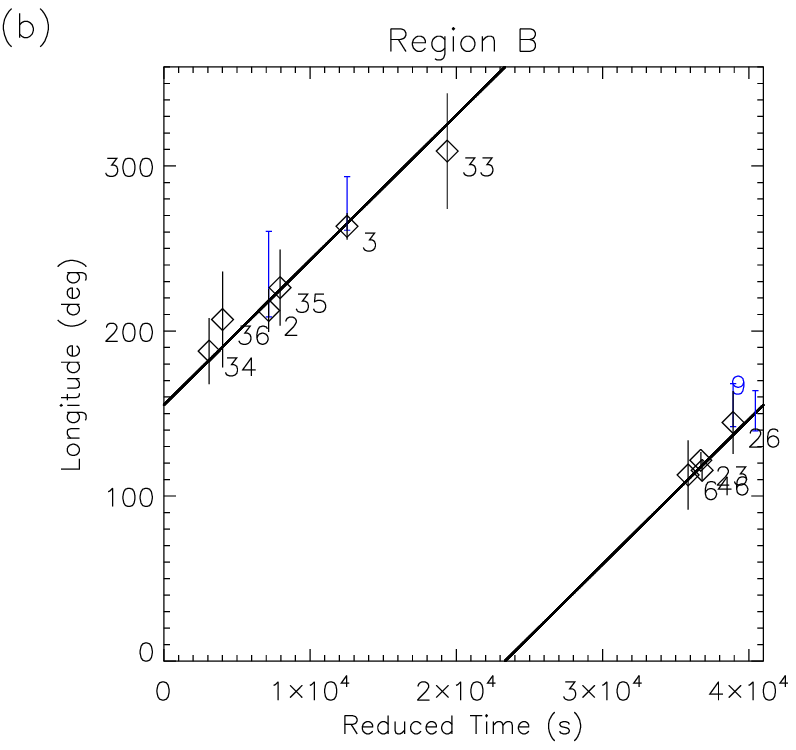}}}
  {\scalebox{1.0}{\includegraphics{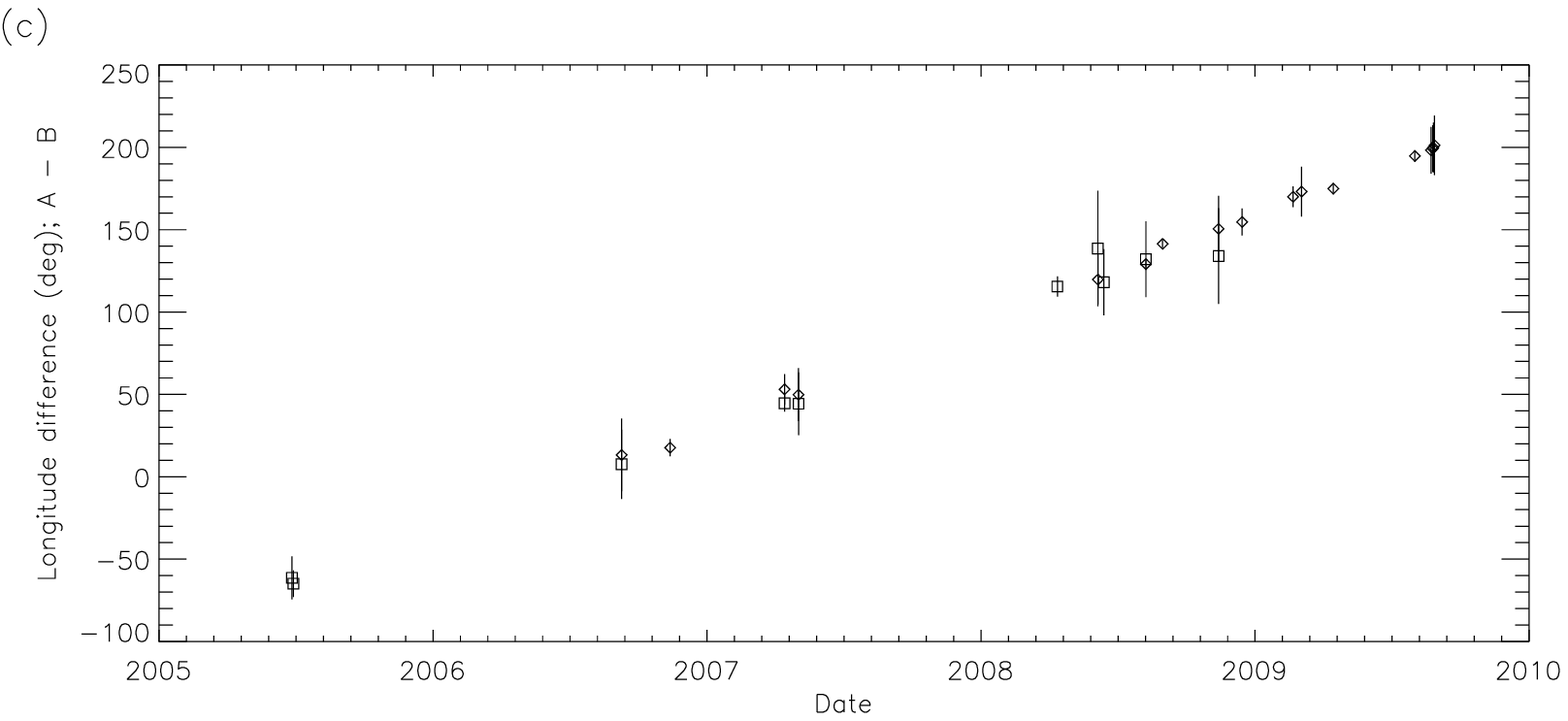}}}
 Fig. \ref{fig:lonplots}
 \end{center}
\end{figure}

\end{document}